\documentclass[letter]{aa}
\usepackage{graphicx}
\usepackage{txfonts}
\usepackage{subcaption}         \usepackage{lscape}             \usepackage{placeins}

\usepackage{hyperref}
\hypersetup{
  colorlinks=true,
  linkcolor=blue,
  citecolor=blue,
  filecolor=magenta,      
  urlcolor=blue,
}

\usepackage{color}

\begin{document}

\title{Hourly radio variability of PDS\,70c  from \\ time-differential photometry\thanks{\email{simon@das.uchile.cl}}
}

\author{
  Simon~Casassus\inst{\ref{UCH},\ref{DO}}
  \and
  Miguel~C\'arcamo\inst{\ref{USACH},\ref{DO}}
  \and
  Oriana~Dom\'inguez-Jamett\inst{\ref{UCH}}
  \and
  Yuhiko~Aoyama\inst{\ref{yuhiko1}}
  \and
  Gabriel-Dominique~Marleau\inst{\ref{UDE},\ref{MPIA}}
  \and
  Ond\v{r}ej~Chrenko\inst{\ref{charles}}
  \and
  Hauyu~Baobab~Liu\inst{\ref{Baobab1},\ref{Baobab2}}
  \and
  Barbara~Ercolano\inst{\ref{LMU}}
}

\institute{Departamento de Astronom\'{\i}a, Universidad de Chile, Casilla 36-D, Santiago, Chile\label{UCH}
  \and
  {Data Observatory Foundation, Eliodoro Y\'a\~{n}ez 2990, Providencia, Santiago, Chile} \label{DO}
  \and
  University of Santiago of Chile (USACH), Faculty of Engineering, Computer Engineering Department, Chile \label{USACH}
  \and
  {School of Physics and Astronomy, Sun Yat-sen University, Guangdong 519082, People's Republic of China} \label{yuhiko1}
  \and
  {Fakult\"at f\"ur Physik, Universit\"at Duisburg-Essen, Lotharstraße 1, 47057 Duisburg, Germany} \label{UDE}
\and
  {Max-Planck-Institut f\"ur Astronomie, K\"onigstuhl 17, 69117 Heidelberg, Germany} \label{MPIA}
  \and
  Charles University, Faculty of Math and Physics, Astronomical Institute, V Hole\v{s}ovi\v{c}k\'{a}ch 747/2, 180 00 Prague 8, Czech Republic\label{charles}  \and
  {Department of Physics, National Sun Yat-Sen University, No.~70, Lien-Hai Road, Kaohsiung City 80424, Taiwan} \label{Baobab1}
  \and 
  {Center of Astronomy and Gravitation, National Taiwan Normal University, Taipei 116, Taiwan} \label{Baobab2}
  \and
  {University Observatory, Faculty of Physics, Ludwig-Maximilians-Universit\"at M\"unchen, Scheinerstr.~1, 81679 Munich, Germany} \label{LMU}
 }

\date{Received 22 February 2026/ Accepted ...}

\abstract
{The  radio emission mechanisms from accreting  protoplanets, and their  variability, link  observations and physical properties.
}
{ We revisit the variability of the $\sim$343\,GHz (ALMA Band 7) flux
  density from PDS\,70c ($F_{\rm B7}$).
} 
{ The subtraction of the extended time-averaged signal may enable the
  measurement of the flux density from variable and embedded point
  sources.  Visibility alignment and self-calibration yields close to
  thermal residuals in each execution block (EB) of ALMA observations, allowing the
  time-differential photometry of point-source  in the
  visibility domain. The variability of PDS\,70c is  checked against synthetic control point sources.     
}
{In images of the 2017 ALMA dataset, with three $\sim$1\,h EBs,
  PDS\,70c was detected only on 6 December 2017, where $F_{\rm B7}$ rose by
  $228\%\pm69\%$ (3.3$\sigma$).  Time-differential photometry confirms
  a rise by $170\%\pm46\%$ (3.7$\sigma$).  An application to  $\sim$2\,h EBs  from the 2023  dataset resulted in constant flux densities, within a scatter of  $\sim$15\%. However, $F_{\rm B7}(t)$ shows some scatter when splitting the deep 2023 EBs in 20\,min intervals, with a $\chi^2$ test significant at 2.6\,$\sigma$, and an intrinsic dispersion of $49\%\pm21\%$. 
}
{The radio variability of PDS\,70c, observed over hours but averaged out on longer  timescales, is indeed expected if the signal is due to H\,{\sc i} free-free from an accretion shock on a circum-planetary disk surface. A planet-to-environment mass ratio  $<10^{-4}$  is required to avoid smoothing by radiative diffusion if the signal is due to thermal emission from the environment. }

\maketitle
\nolinenumbers

\section{Introduction}

Most embedded protoplanet candidates known to date, such as
PDS\,70b,c, AB\,Aur\,b, WISPIT\,2b and 2MASS\,J16120668-3010270\,b,
are H\,$\alpha$ sources and thought to be undergoing mass accretion
\citep[][]{Keppler2018A&A...617A..44K, Haffert2019NatAs...3..749H,
  Wang2021AJ....161..148W, Christiaens2024A&A...685L...1C,
  Hammond2025MNRAS.539.1613H, Currie2025ApJ...990L..42C,
  vanCapelleveen2025ApJ...990L...8V, Li2025ApJ...990L..70L,
  Close2025ApJ...990L...9C}. Among these candidates, only PDS\,70c is
known to have a radio counterpart
\citep[][]{Isella2019ApJ...879L..25I, Benisty2021ApJ...916L...2B,
  Fasano2025A&A...699A.373F, Dominguez2025AA...702A..18D}. The sub-mm
continuum signal from PDS\,70c is particularly important as it
constrains the planetary accretion rate and conditions in its
environment, provided the radio emission mechanisms are understood. By
contrast, H$\alpha$ observations \citep[][]{Zhou2025ApJ...980L..39Z,
  Close2025AJ....169...35C} can be affected by variable extinction
\citep[][]{Cugno2025AJ....170..317C}.

Based on nearly-coeval multi-frequency ALMA measurements,
\citet[][]{Dominguez2025AA...702A..18D} suggested that, contrary to
the standard hypothesis, the sub-mm signal is probably not due to
thermal dust emission, as its optically thick spectral index over
97\,GHz, 145\,GHz, and 343\,GHz (Band\,7,
hereafter IB23) is difficult to reconcile with the partially thick
emission from a circumplanetary disk or envelope. It may instead be
due to thermal plasma radiation, either with residual ionization, as
in a cool star (i.e. due to the H$^-$ or H$_{2}^{-}$ free-free
continua), or fully ionized, as in an H\,{\sc ii} region. A 671\,GHz drop suggests that the most likely explanation of the signal
from PDS\,70c is H\,{\sc i} free-free from a thin H\,{\sc ii} region
at the shocked surface of a circum-planetary disk (CPD) undergoing
mass-loading \citep[as described by ][]{Aoyama2018ApJ...866...84A,
  Szulagyi2020ApJ...902..126S}.

The spectral analysis of \citet[][]{Dominguez2025AA...702A..18D}
rests on the hypothesis that variability is negligible, which is
justified by the short time-span encompassing the multi-frequency
observations, of at most two months. Indeed, the IB23 dataset,
acquired in nine execution blocks (EBs) spanning six months, suggests
that the Band\,7 flux density of PDS\,70c (hereafter $F_{\rm B7}$) is
constant within a standard deviation of 10\%. In fact, $F_{\rm B7}$
appears constant (Table\,\ref{table:PDS70cprev}) when comparing the
discovery observation in the 2017 dataset \citep[][]{Isella2019ApJ...879L..25I}, hereafter IB17 \citep[following the nomenclature of ][]{Benisty2021ApJ...916L...2B},
with measurements in 2019 \citep[hereafter
LB19,][]{Benisty2021ApJ...916L...2B}, in 2021
\citep[][]{Fasano2025A&A...699A.373F}, and in 2023
\citep[i.e. in  IB23,][]{Dominguez2025AA...702A..18D}.

However, motivated by questions on the uncertainties in restored
images and in the concatenation of observations,
\citet[][]{CasassusCarcamo2022MNRAS.513.5790C} and
\citet{Casassus2023MNRAS.526.1545C} chose PDS\,70 as an example source
to test an original scheme for multi-epoch visibility alignment (the
{\tt VisAlign} package), and apply automatic self-calibration (with
the {\tt snow} package). Surprisingly, their revisit of IB17 yielded the non-detection of
PDS\,70c, despite sufficient signal-to-noise ratio (S/N) to reproduce the
result of \citet[][]{Isella2019ApJ...879L..25I}. This non-detection led to the
conclusion for variability, of at least $\sim$40\% over $\sim$2\,yr.

We revised the analysis of
\citet[][]{CasassusCarcamo2022MNRAS.513.5790C} with the goal to
resolve the discrepancy with all other measurements of $F_{\rm
  B7}$. It turns out that IB17 is composed of three EBs, dated
2, 3, and 6 December. However, perhaps as a result of a
bug in the handling of {\tt CASA} tasks (which we attribute to
intensive calls to the {\tt mstransform} task, combining both time and
channel averaging), the EB from 6 December was missing. As shown below, the inclusion
of 6 December brings $F_{\rm B7}$ in agreement with the original report by
\citet[][]{Isella2019ApJ...879L..25I}. The variability reported by
\citet[][]{CasassusCarcamo2022MNRAS.513.5790C} therefore applies to
timescales of days or less, rather than years.

Here, we report on a technique to estimate the flux density of
unresolved, but variable, embedded sources, and apply it to ALMA Band\,7
datasets of PDS\,70 (Sect.\,\ref{sec:diffphot}, the datasets are
summarised in Table\,\ref{table:PDS70cprev}). We then discuss the
expected timescales of the radio variability of PDS\,70c for relevant
emission mechanisms (Sect.\,\ref{sec:disc}).  We conclude for a
tentative detection of hourly variability, which favours an accretion
shock at the CPD surface as the origin of the signal
(Sect.\,\ref{sec:conc}).

\section{Analysis and results} \label{sec:diffphot}

\begin{table}
\caption{\label{table:PDS70cprev} Summary of the datasets used here and previous measurements for $F_{\rm B7}$, the Band\,7 flux density for PDS\,70c.} 
\centering
\begin{tabular}{lccc}
  \hline\hline
  & $\theta$ & $F_{\rm B7}$                        & Year \\
  & mas & $\mu$Jy &     \\ \hline
  IB17\tablefootmark{a} &   $\sim$60 & $106\pm19$\tablefootmark{a} & 2017\\
  LB19\tablefootmark{b} &   $\sim$50 & $118\pm24$\tablefootmark{c} & 2019\\
  IB23\tablefootmark{d} &   $\sim$60 & $121\pm13$\tablefootmark{d} & 2023\\
\hline
\end{tabular}
\tablefoot{
\tablefoottext{a}{\citet{Isella2019ApJ...879L..25I}.}
\tablefoottext{b}{\citet{Benisty2021ApJ...916L...2B}.}
\tablefoottext{c}{\citet{Casassus2023MNRAS.526.1545C}.}
\tablefoottext{d}{\citet{Dominguez2025AA...702A..18D}.}
}
\end{table}

\subsection{Individual EB imaging of IB17}

\begin{figure*}
  \centering
\includegraphics[width=0.9\textwidth]{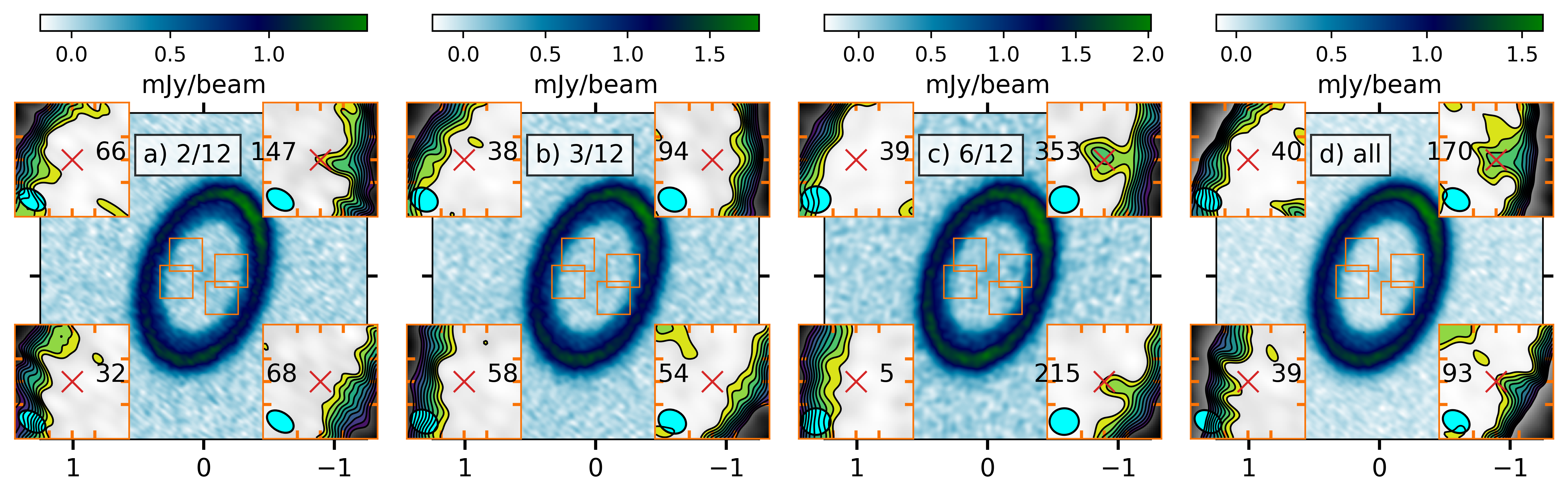}
      \caption{Restored images for IB17, with Briggs parameter $r=0.3$. Insets zoom on PDS\,70c, to the north-east,  and symmetric positions along disk axes. Image noise and beams  are provided in Table\,\ref{table:IB17}. Peak intensities are all equal when brought to a common beam.
        \label{fig:daily}}
\end{figure*}

Details on the alignment and self-calibration procedure are given in
Appendix\,\ref{sec:IB17}. Imaging of the whole IB17 dataset
(Fig.\,\ref{fig:daily}d) confirms the detection of PDS\,70c by
\citet{Isella2019ApJ...879L..25I}. Its flux density, given by the peak
intensity at the position of the point source in restored maps, is
$\langle F_{\rm B7}\rangle = 172\pm30$\,$\mu$Jy, where the average
represents a joint measurement over all 3 EBs.  As detailed in
Appendix\,\ref{sec:IB17} alternative measurements with Gaussian fits
yield similar values, within the errors (Table\,\ref{table:IB17}).
$\langle F_{\rm B7} \rangle$ is consistent, at 1.8$\sigma$, with the
value reported by \citet{Isella2019ApJ...879L..25I}.

However, imaging of individual EBs, shown in Fig.\,\ref{fig:daily},
revealed no signal under PDS\,70c in 2 and 3 December. Since
the source is known to exist, rather than assigning zero flux density
we reported, in Table\,\ref{table:IB17}, the specific intensity at the
expected position. This strategy for point-source photometry
incorporates the prior knowledge of the source position \citep[as
discussed by ][ their sect.\,2.3]{Dominguez2025AA...702A..18D}, and
is sometimes referred to as ``forced photometry''
\citep[][]{Samland2017A&A...603A..57S}.

The source is, however, quite conspicuous in 6 December, at
$370\pm\,65\,\mu$Jy or $\sim$5.7\,$\sigma$ (as given by the peak
specific intensity).  The $F_{\rm B7}$ measurement on 6 December is roughly
twice the average value of $\langle F_{\rm B7}\rangle$, and the
difference between 3  and 6 December  is $276\pm82.6$\,$\mu$Jy and
significant at 3.3$\,\sigma$. Fixing  the average value of $F_{\rm B7}$ as given by \citet[][Table\,\ref{table:PDS70cprev}]{Dominguez2025AA...702A..18D}, this rise corresponds to   $228\%\pm69\%$.

\subsection{Time-differential photometry}  \label{sec:uvplane}

The preceding estimates for $F_{\rm B7}$ are, for simplicity, directly
read out as the peak intensity under PDS\,70c in restored images,
where the source is  separated from surrounding
signal. But this approach requires giving artificially
more weight to the longer baselines, overriding error propagation at
the expense of sensitivity. In other words, with natural weights the
uncertainties would be lower, but the source would no longer be
isolated from the outer ring (Fig.\,\ref{fig:daily}), thus
introducing systematic errors.

We  also  tested the time variable nature of the radio signal from
PDS\,70c by measuring the relative variations in $F_{\rm B7}$ in the $uv$ plane. Our method 
is to fit a point source, fixed at the location of PDS\,70c
($\vec{x_\circ}$, obtained with an elliptical Gaussian fit to 6 December),
in the visibility residuals $V^R$, after subtraction of the model
visibilities calculated from the model image that best fits the
concatenated dataset. Least-squares minimisation yields
\citep[notation follows from ][their
appendix]{CasassusCarcamo2022MNRAS.513.5790C}:
\begin{equation}
  F_\nu  = \frac{\sum_{l=1}^{L} \omega_{l}~ \Re\left[  V^R_{l} e^{-2\pi i \vec{u}_{l}\cdot\vec{x_\circ}}\right]  }{ \mathcal{A}(\vec{x_\circ})  \sum_{l=1}^{L} \omega_{l} } \label{eq:PS}
\end{equation}
where $\vec{u}_{l}$ is the $uv$-plane position for datum $V^R_{l}$ and $\mathcal{A}(\vec{x_\circ})$ is the primary-beam attenuation. The associated errors are
\begin{equation}
  \sigma(F_\nu) = \frac{1}{ \mathcal{A}(\vec{x_\circ})  \sqrt{\sum_{l=1}^{L} \omega_{l} }  }. \label{eq:sigmaF}
\end{equation}
A similar measurement of $F_{\rm B7}$ can also be obtained from the
gridded visibilities. As shown in Appendix~\ref{sec:griddedphot}, the
gridded approach bypasses the model image, but it collapses the
spectral information, and depends on the gridding scheme. Since the
model image is used anyway in the self-calibration procedure, we opted
for photometric extraction on the raw data. Both approaches yield
similar results for a flat spectral index (Fig.\,\ref{fig:residuals}).

Such time-differential photometry is deceivingly simple, as it assumes
perfect calibration. Any imperfection yields non-thermal residuals. For instance, omitting EB alignment with {\tt
  VisAlign}, or omitting the amplitude self-calibration, will yield
such residuals, as documented in Appendix\,\ref{sec:residuals} and
Fig.\,\ref{fig:residuals}. However, alignment and self-calibration
allow reaching close to thermal residuals, even in natural weights.

Another difficulty with the point-source photometry is that the
uncertainties in Eq.\,\ref{eq:sigmaF} derive  from the
visibility weights, whose absolute calibration is uncertain. This
can be overcome with the
replacement of the original weights by the root-mean-square dispersion
of each visibility datum, taken over all individual integrations, as
performed with the {\tt CASA} task {\tt statwt}. An additional imaging run
is in order after the application {\tt statwt}, to check that it did
not appreciably deteriorate  dynamic range. 

A complementary method to estimate the uncertainty on $uv$-plane
point-source photometry is to take statistics on flux-densities
extracted at randomly distributed positions. We refer to this estimate
as the ``spatial scatter''.  At a given epoch we applied
Eq.\,\ref{eq:PS} at  $N = 120$ random positions, and took
the dispersion of the resulting $N$ flux densities as a measure of its
uncertainty. In the presence of extended and strong signal, this
uncertainty will depend on position since imperfect antenna gains are
multiplicative. Given that the source in consideration is
approximately a simple ring, we chose to place the random positions
along a projected circular orbit matching that of PDS\,70c (but
excluding the position of PDS\,70c within one natural-weight beam).

\begin{figure}
  \centering
\includegraphics[width=0.9\columnwidth]{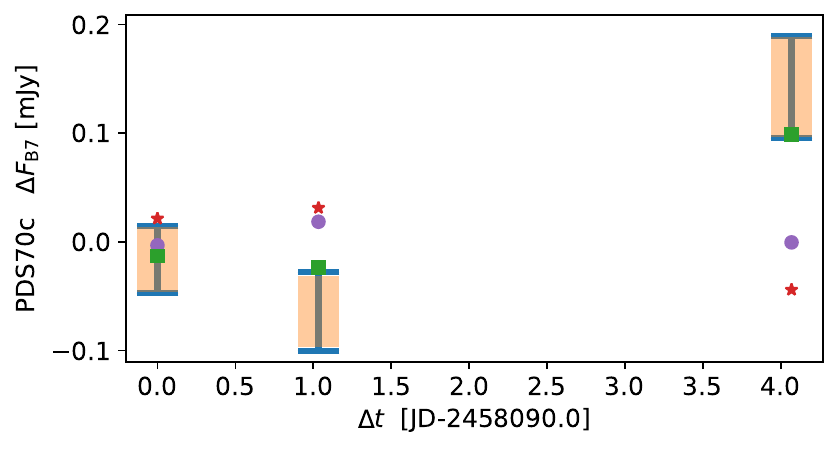}
  \caption{Time-differential photometry of PDS\,70c. The blue $\pm1\,\sigma$ error bars (2$\,\sigma$ in total)  record  $\Delta F_{\rm B7} =  F_{\rm B7}-\langle F_{\rm B7}\rangle$ as a function of Julian day for each EB. The same quantity extracted from the spatial scatter  (see  text)  is shown in thick orange error bars.  Each type of  marker corresponds to 3 synthetic and constant point-sources, injected at the  positions of the insets in Fig.\,\ref{fig:daily}. 
    \label{fig:Fvar}}
\end{figure}

The close to thermal residuals justify the use of Eq.\,\ref{eq:PS} to
extract point-source flux-densities in the visibility residuals, and
thus measure
$\Delta F_{\rm B7}(t) = F_{\rm B7}(t) - \langle F_{\rm B7}\rangle
$. The result is summarised in Fig.\,\ref{fig:Fvar}, where we see that
$F_{\rm B7}$ is consistent between 2  and 3 December, but rises by
$206\pm56\,\mu$Jy from 3 to 6 December, which is significant at
3.7$\,\sigma$. Fixing $F_{\rm B7}$ to the value for IB23 in
Table\,\ref{table:PDS70cprev}, this rise corresponds to $170\%\pm46\%$. The
  quoted errors are derived from the spatial scatter, and are very
  close to the errors given by the visibility weights (after the
  application of {\tt statwt}).

As a test on the significance of the variability of PDS\,70c, we
injected three synthetic and constant point-sources in the
concatenated dataset, at symmetric positions from PDS\,70c relative to
the disk minor and major axis. These positions match the centers of
the insets in Fig.\,~\ref{fig:daily}. Using the uncertainties inferred
from the spatial scatter, as for PDS\,70c, the largest variation in
$\Delta F_{\rm B7}(t)$ for the synthetic point-sources is
$122\pm 56$\,$\mu$Jy, or 2.2\,$\sigma$.

An application of differential photometry to other Band\,7 datasets is
documented in Appendix\,\ref{sec:IB23}. Neither LB19 nor IB23 present
detectable variability among EBs.

It is intriguing that the daily variability of $F_{\rm B7}$ is picked
up in IB17, but not in IB23. The deep IB23
dataset spanned six\,months in nine EBs with integrations of over
2\,h, while the IB17 EBs are 1\,h long. Yet a search for variability,
using a related approach to the time-differential photometry presented
here (but subtracting a Gaussian fit to PDS\,70c in the model
  image, so the source stands out in the residuals), yielded a
constant value for $F_{\rm B7}$, with a scatter of $\sim$10\% among
EBs \citep[][]{Dominguez2025AA...702A..18D}. Perhaps the source underwent a peculiar glitch right on
6 December.  However, given that the IB23 EBs are more than twice
longer than in IB17, an interesting possibility is that the
variability of PDS\,70c occurs on timescales of minutes or hours
rather than days, and that it is averaged out in longer integrations.

The difficulty in observing fine-grained time variability is that
splitting into shorter integrations results in noisier
measurements. As documented in Appendix\,\ref{sec:IB23}, we nonetheless pick up  some
scatter in $\Delta F_{\rm B7}(t)$ in $\sim$20\,min intervals of the
IB23 dataset, corresponding to a reduced $\chi^2$ of 1.7 with 41
degrees of freedom, or significant at 2.6\,$\sigma$. The intrinsic
rms variability of PDS\,70c is $\sigma_{\rm int} = 59\pm25\,\mu$Jy.

\section{Discussion} \label{sec:disc}

The constraints on the radio variability of PDS\,70c bring information
on the emission mechanisms in the circum-planetary environment.  For
example, radiative heating from accretion bursts could provide
external forcing to the emitting medium. The radio variability would
result from the intrinsic variability of accretion smoothed by the
timescale associated to thermal diffusion.

The interpretation that $F_{\rm B7}$ stems from an accretion shock at
a CPD surface around PDS\,70c \citep[][]{Dominguez2025AA...702A..18D}
naturally accounts for fast variability. Since the accretion flows
within the Hill sphere can be approximated as free fall
\citep[e.g.][]{Tanigawa2012ApJ...747...47T, Fung2016ApJ...832..105F},
the free-free emission responds almost directly to fluctuation of
inflow from the protoplanetary disk, with only a delay corresponding
to the free-fall time. Such inflows can vary by factors of a few along
the planetary orbit in turbulent disks
\citep[][]{Gressel2013ApJ...779...59G} or for eccentric planets
\citep[][]{Li2023MNRAS.526.5346L}, so the accretion-shock density and
thus the free-free emission flux can vary by similar factors. However,
in their hydro-dynamical models of the CPD surface shock,
\citet{Takasao2021ApJ...921...10T} find that the interaction between
the infalling gas and the CPD triggers high-frequency and stochastic
variability through a thermal instability of the standing shock. The
net mass accretion rate on the CPD surface, and eventually on the
protoplanet itself, result from the interaction of post-shock
cooling and flow dynamics in the CPD. This accretion rate fluctuates
by up to an order of magnitude on timescales much shorter than the
orbital period, and down to hours. By contrast, accretion occurs on a
quasi-steady state if averaged over days, as in the proposal of
Sect.\,\ref{sec:uvplane} to explain the lack of detectable variability
on longer timescales.

In the case of a surface shock, the H\,\,{\sc i } free-free emission
reflects the temperature profile across the shock
\citep[][]{Aoyama2018ApJ...866...84A}.  From the  shock profiles
shown in \citet[][]{Dominguez2025AA...702A..18D} for two
representative pre-shock conditions, in the region where
$\tau(300\,{\rm GHz}) \sim 1$, $T \sim 3\,10^4$\,K to $3\,10^5$\,K, at
depths of $\sim 10^4$\,cm and $\sim 10^5$\,cm, respectively. Given
that the main source of opacity is optically thin, except at radio
frequencies, energy transfer occurs over a light-crossing time, which
is essentially instantaneous.

Alternatively the radio signal may stem from the bulk mass in the
planetary environment, under external forcing.  Assuming energy
transfer in an optically thick medium via Rosseland diffusion, the
typical photon mean-free-path is $l_R = 1/ (\kappa_R \, \rho)$, where
$\kappa_R$ is the Rosseland opacity and $\rho$ is the mass density.
The diffusion timescale is $\tau_d = N_{\rm scat} \, l_R / c $, with
$N_{\rm scat} = (L / l_R)^2$, and where $L$ is the typical size of the
emitting region.

We distributed the mass in the planetary environment $M_{\rm env}$, in
a cubic box with side $L$, representing the linear size of a CPD or of
an envelope. The outer radius of the CPD could be 1/3 of the Hill
radius, or $\sim$1\,au for PDS\,70c, and a typical aspect ratio is
$\sim 0.1$, so $L \sim 0.1$\,au. Then the diffusion timescale is
$\tau_d = M_{\rm env} \,\kappa_R / (c \, L)$. Reducing the box side,
or increasing $\kappa_R$, results in larger diffusion timescales for a
given mass.

The floor $\kappa_R$ in a dust-deprived medium with residual ionisation, is due to
the H$^-$ or H$_2^-$ free-free continua and to atomic or molecular
lines \citep[e.g.][]{Malygin2014A&A...568A..91M}. In a CPD accounting
for the PDS\,70c radio spectrum
\citep[][]{Dominguez2025AA...702A..18D}, the annulus with the peak
contribution to the net flux density, at $R=0.05 \, \rm au$, has
mid-plane $\rho \sim 10^{-8}\, {\rm g \,cm}^{-3}$ and
$T \sim 2000 \, \rm K$, and $\kappa_{\rm R} \gtrsim 0.1$ \citep[for
solar metallicity,][their Fig.\,3]{Malygin2014A&A...568A..91M}. The
contribution from dust to the floor $\kappa_R$ is negligible, at least
for large dust grains, since $\kappa_R \sim 0.001 \,$cm$^2$\,g$^{-1}$
for a standard gas-to-dust mass ratio of 100
\citep[e.g.][]{Chrenko2025A&A...700A..82C}.

An environment mass of $M_{\rm env} \sim 0.07 M_{\rm Jup}$, for a
planet-to-CPD mass ratio of $10^{-2}$ \citep[considering a planet mass
of $7\,M_{\rm Jup}$][]{Wang2021AJ....161..148W}, is in practice an
upper limit. The corresponding dust mass, of $0.2\,M_{\oplus}$, is
larger than the total mass of the Galilean moons, which is
$0.07\,M_{\oplus}$. It is also much larger than the dust mass inferred
from modelling $F_{\rm B7}$ assuming thermal dust emission, or
$\sim 1.4\times 10^{-2}\,M_{\oplus}$
\citep[][]{Shibaike2024A&A...687A.166S}.

If $M_{\rm env} = 0.07 M_{\rm Jup}$, $L = 0.1$\,au, and
$\kappa_R = 0.1\,$cm$^2$\,g$^{-1}$, then $\tau_d = 3.5\,{\rm d}$.
Fixing $\tau_d = 1$\,h would require a box with side
$L = \kappa_R \times M_{\rm env} / (c \times \tau_d) \sim 8$\,au,
which is much larger than the constraint $L <1$\,au (from
the limit on the angular extent of the source). Alternatively, the
environment mass would have to be reduced to
$M_{\rm env} = 8\times 10^{-4} M_{\rm Jup}$ to match an hourly
variability.

\section{Conclusions} \label{sec:conc}

Imaging of individual ($\sim$1\,h) execution blocks (EBs) in the 2017
Band\,7 observations of PDS\,70 shows that PDS\,70c is conspicuous on
6 December 2017, but undetectable in both 2 and 3 December. Estimates of the
flux density from PDS\,70c in the restored images ($F_{\rm B7}$)
result in a rise from 3  to 6 December, significant at 3.3$\,\sigma$.
 
Time-differential photometry allows the extraction of the relative
variations of $F_{\rm B7}$ in the $uv$-plane, at higher significance
than in the restored images, where the source is
confused with the outer ring.  The residual Band\,7 flux density of
PDS\,70c ($\Delta F_{\rm B7}$, after subtraction of the time-averaged
extended emission, also rose from 3 to 6 December, by
3.7$\,\sigma$. Since the residuals in single EBs are not perfectly
thermal, this estimate is tentative.

Tests on the variability of PDS\,70c in Band\,7 observations from
2023, with twice longer EBs than in 2017, yielded a limit of
$\sim$15\%, as given by the dispersion of $\Delta F_{\rm B7}$ among seven
EBs and consistent with a constant. However, splitting these 2\,h EBs
in 20\,min intervals resulted in some intrinsic variability,
significant at 2.6\,$\sigma$, and with an intrinsic rms scatter of
$59\pm25\,\mu$Jy. These results tentatively suggest that the radio
signal from PDS\,70c is variable by $\sim 50\%\pm 21\%$ on timescales
of hours, but that its variability is averaged out on longer timescales.

As shown recently, existing models of surface accretion onto a CPD
account for the PDS\,70c radio spectrum
\citep[][]{Dominguez2025AA...702A..18D}. Such models also predict
high-frequency and stochastic planetary accretion variability, on
timescales of hours, and averaged to a steady state on longer
timescales \citep{Takasao2021ApJ...921...10T}. If the radio signal
stems from the surface shock itself, then the observed radio
variability directly follows the accretion variability. In turn, if
the radio signal stems from bulk mass in the planetary environment,
then the environment mass must be at least a factor $\sim 10^{-4}$
lower than the planetary mass to bring the diffusion timescale down to
1\,h.

\begin{acknowledgements}
  We thank the referee, Luis Pe\~na-Mo\~nino, for a constructive
  report, and also Myriam Benisty and Daniele Fasano for sharing their
  version of the IB17 dataset, which included Dec. 6.  S.C. and
  M.C. acknowledge support from Agencia Nacional de Investigaci\'on y
  Desarrollo de Chile (ANID) given by FONDECYT Regular grant 1251456,
  and ANID project Data Observatory Foundation
  DO210001. G.-D.M. acknowledges the support of the Deutsche
  Forschungsgemeinschaft (DFG) through grant MA~9185/2-1.
  O.C. acknowledges the Czech Science Foundation (grant 25-16507S),
  the Charles University Research Centre program
  (No. UNCE/24/SCI/005), and the Ministry of Education, Youth and
  Sports  through the e-INFRA CZ (ID:90254).
\end{acknowledgements}

\bibliographystyle{aa_url} \bibliography{refs}

\begin{appendix}

\section{Alignment, self-calibration and image-plane $F_{\rm B7}$ measurements in the  IB17 dataset} \label{sec:IB17}

Description of the IB17 ALMA dataset, alignment of EBs and imaging follow
from \citet{CasassusCarcamo2022MNRAS.513.5790C},
\cite{Casassus2023MNRAS.526.1545C}, and references therein. In brief,
the 3 continuum spectral windows were time-averaged to 6.06\,s, and
binned into 8 channels.  Individual EBs from 3  and 6 December 2017
were then aligned to the first EB, from 2 December, using the {\tt
  VisAlign} package, including a correction for visibility amplitudes
$\alpha_R$. The resulting offsets were $\alpha_R = 1.102\pm0.002 $,
$\Delta\alpha = 0\farcs00845 \pm 0\farcs00034$,
$\Delta\delta = -0.00045^{+0.0004}_{-0.00044}$, for 3 December, and
$\alpha_R = 1.225\pm 0.003$,
$\Delta\alpha = 0\farcs0107\pm0\farcs0004$,
$\Delta\delta = 0\farcs0043\pm 0.0005$, for 6 December. An offset in
flux scale of $\sim$20\% is consistent with the nominal calibration
uncertainty in ALMA Band\,7, of $\sim10\%$ rms, while the positional
offsets along R.A. and Dec. are all consistent with the nominal
pointing accuracy, of $\sim 1/10$ the naturally-weighted beam of
$0\farcs09\times 0\farcs08$.

Following alignment, we concatenated the data into a single
measurement set, and applied the {\tt snow} package for automatic self
calibration \citep[{\tt snow} replaces the tclean imager by {\tt
  gpu-uvmem},][in otherwise standard {\tt CASA}
self-calibration]{Carcamo2018A&C....22...16C}. Phase-only
self-calibration did not yield improvements. However, a single
iteration in amplitude and phase self-calibration, with a solution
interval limited to a single scan (roughly 50\,s), improved the
peak S/N from 55 to 115.

Once self-calibrated, each EB was split for separate imaging with {\tt
  gpu-uvmem}, in pure-$\chi^2$ reconstructions (i.e. without entropy
regularization, but truncating the optimization at 10 iterations). The
(non-parametric) model images were restored using Briggs weights with
a robustness parameter $r=0.3$. The resulting images are shown in
Fig.\,\ref{fig:daily}.

Table\,\ref{table:IB17} provides a summary of the image properties in
the IB17 dataset. We also tabulate the values of $F_{\rm B7}$ measured
in each EB as well as in the concatenated dataset. The elliptical
Gaussian fits were carried out following two strategies: 1- by fitting
a beam, assuming that the source is unresolved, and 2- by fitting an
elliptical Gaussian.  Fitting a beam resulted in a bias for larger
flux densities to account for extended low-level emission. Fitting an
elliptical Gaussian was sensitive on the definition of the zoomed
region around PDS\,70c, and turned out to be impossible in the case of
the concatenated dataset, where the source is connected to the ring.

\begin{table*}
\caption{\label{table:IB17} Summary of image properties and flux densities for PDS\,70c ($F_{\rm B7}$) in the  IB17 dataset, as obtained  in the sky images.} 
\centering
\begin{tabular}{lcccccc}
  \hline\hline
Date   & Beam\tablefoottext{a} & Noise\tablefoottext{b} & Peak\tablefoottext{c} & Beam fit\tablefoottext{d} & \multicolumn{2}{c}{Gauss fit\tablefoottext{e} }  \\ 
  & mas                    & $\mu$Jy                & $\mu$Jy              &  $\mu$Jy                     &   $\mu$Jy       &  mas \\ \hline
  2/12/2017 &   $64 \times 41$ &  44  & $147\pm44$\tablefoottext{f}  &  $\cdots$    &    $\cdots$   & $\cdots$    \\
  3/12/2017 &   $63 \times 50$ & 51 & $94\pm51$\tablefoottext{f}  &  $\cdots$    &    $\cdots$     & $\cdots$  \\
  6/12/2017 &   $64 \times 58$ & 65 & $370\pm65$  & $502\pm65$  & $878\pm210$ &  $111^{+17}_{-10} \times 79^{+12}_{-12}$     \\
  All &    $62 \times 46$ & 30 &$172\pm30$   & $250\pm30$ &  $\cdots$ $\tablefoottext{g}$ & $\cdots$ $\tablefoottext{g}$  \\ \hline \end{tabular}
\tablefoot{
\tablefoottext{a}{Elliptical Gaussian fit to the beam.}
\tablefoottext{b}{Standard dispersion of the residual images (i.e. 1$\sigma$ noise).}
\tablefoottext{c}{$F_{\rm B7}$ from peak intensity under PDS\,70c in restored maps.}
\tablefoottext{d}{$F_{\rm B7}$ from fitting the elliptical Gaussian clean beam.}
\tablefoottext{e}{$F_{\rm B7}$ from fitting an elliptical Gaussian. The best fit Gaussian full-width-at-half-maximum along major and minor axis are given in the format BMAJ$\times$BMIN.}
\tablefoottext{f}{Non-detection, we report the intensity at the expected position of PDS\,70c.}
\tablefoottext{g}{An elliptical Gaussian fit is not possible because the source is confused with the bright ring.}
}
\end{table*}

\section{Residual maps on single EBs and impact of  alignment  and self-calibration} \label{sec:residuals}

\begin{figure}
  \centering  
\includegraphics[width=0.95\columnwidth]{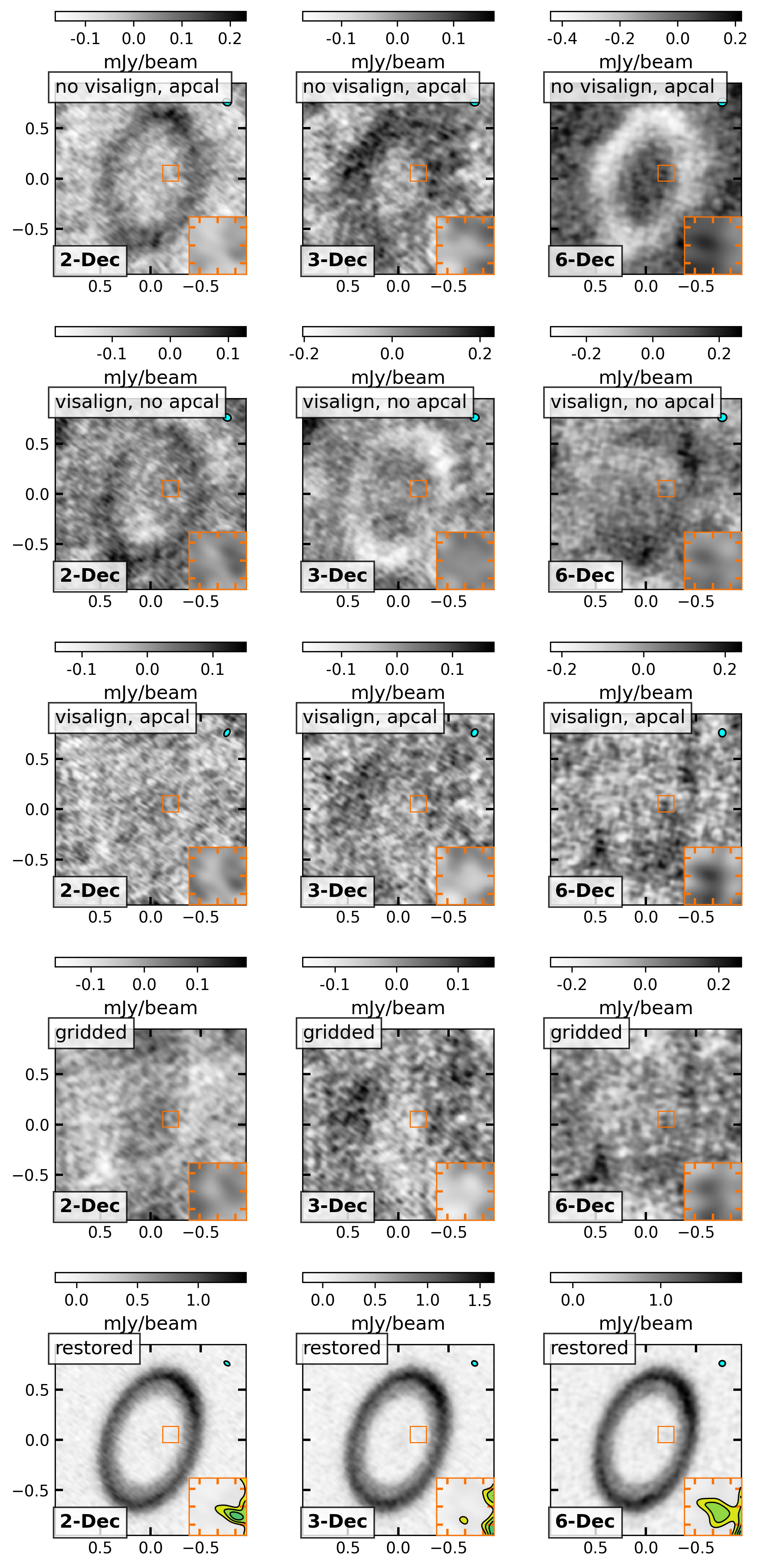}
  \caption{
Dirty map residuals in natural weights for each EB, after subtraction of the model visibilities from the  concatenated dataset. Insets zoom on PDS\,70c, in the same colour stretch as for the image. Each row, from top to bottom,  shows: ``no visalign, apcal'', without  {\tt visalign} alignment, but after amplitude and phase self-calibration (apcal);  ``visalign, no apcal'': with alignment, but no self-calibration; ``visaligned,  ap'': here with both alignment and self-calibration, and reaching close to thermal residuals; ``gridded'': also with both alignment and self-calibration, but from the gridded approach (i.e. $I_{\rm diff}$ in Eq.\,\ref{eq:Idiff}); ``restored'': restored maps, obtained with $r=0.3$ (as in Fig.\,\ref{fig:daily}).
    \label{fig:residuals}}
\end{figure}

Reaching thermal residuals in a dataset comprised of several EBs is
 routinely achieved, provided with sufficiently accurate
calibration. However, because of the varying flux scale and pointing,
the residuals under individual EBs will, in general, not be
thermal. The application of both EB alignment (with {\tt VisAlign}),
along with amplitude and phase self-calibration, allowed reaching
close to thermal residuals in the 2017 dataset, as illustrated in
Fig.\,\ref{fig:residuals}. The omission of alignment or
self-calibration leads to strong residuals, due to imperfect
calibration of the visibility data.

Amplitude self-calibration of the IB17 dataset required short solution
intervals, down to a single scan or 50\,s. As a result, the derived
antenna gains can be fairly noisy. We found the resulting gains vary
strongly, up to $\sim$100\% rms. These strongly variable gains
modulate the point-source flux densities, in ways that cannot be
corrected for with a single scaling factor (as with {\tt
  VisAlign}). These noisy gains are probably at the origin of the
persistent non-thermal residuals. Improvements in self-calibrations
techniques, such as with ``closure imaging''
\citep[][]{Chael2018ApJ...857...23C}, might allow purely thermal
residuals in this dataset.

\section{Gridded  $uv$-plane time-differential photometry} \label{sec:griddedphot}

The photometric extraction of the variable signal can also be measured
in gridded visibility data.  We can compute the gridded visibilities,
in natural weights, for both the concatenated dataset,
$\tilde{V}_{\rm conc}$, with weights $W_{\rm conc}$, and in each
individual EB, $\tilde{V}_{\rm EB}$, with weights $W_{\rm EB}$. In an
extension of the scheme used in the {\tt visalign} package
\citep[]{CasassusCarcamo2022MNRAS.513.5790C}, the residuals are
\begin{equation}
  \tilde{V}_{\rm diff}  = \tilde{V}_{\rm EB} - \tilde{V}_{\rm conc}   \label{eq:vdiff}
\end{equation}
with weights
\begin{equation}
  W_{\rm diff}  = \frac{W_{\rm conc} \times W_{\rm EB}}{W_{\rm conc} + W_{\rm EB}}.
\end{equation}
Such weights naturally match the $uv$ sampling functions of both
datasets in comparison.

The dirty images of the variable signal in each EB can be used as a
consistency test between the photometry in the raw and gridded
approaches. The point-source flux density in Eq.\,\ref{eq:PS} is a
function of position, and can also be viewed as the dirty map of the
residual visibilities corrected for primary-beam attenuation. In the
gridded approach, the dirty map is 
\begin{equation}
I_{\rm diff}(\vec{x}) = \frac{\sum_{i,j}^{N}  W_{\rm diff}(\vec{u}_{i,j})  ~  \tilde{V}_{\rm diff}(\vec{u}_{i,j}) ~e^{-2\pi i \vec{u}_{i,j}\cdot\vec{x}}} { \sum_{i,j}^{N}  W_{\rm diff}(\vec{u}_{i,j}) }, \label{eq:Idiff}
\end{equation}
in units of Jy\,beam$^{-1}$, and for a grid of side $N$.

Fig.\,\ref{fig:residuals} compares the dirty maps of the best
visibility residuals, i.e. after both alignment and self-calibration,
for both approaches. The row labelled ``{\tt visalign}, apcal'' refers
to the dirty maps of the raw residuals, after subtraction of the model
visibilities that fit the concatenated dataset, while the row labelled
``gridded'' refers to the direct subtraction of the concatenated
dataset, i.e. $\tilde{V}_{\rm difff}$. Both version yield similar
images, with differences below the noise level, which satisfies this
consistency test.

\section{Application to  deep Band\,7 data} \label{sec:IB23}

\begin{figure}
  \centering
\includegraphics[width=0.8\columnwidth]{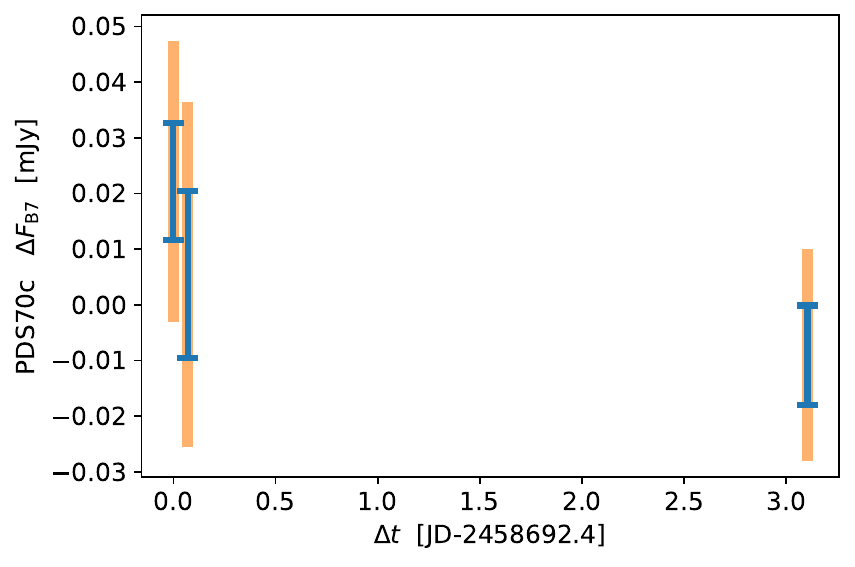}
\caption{Same as Fig.\,\ref{fig:Fvar}, but for the LB19  dataset. The blue error bars stem from the
  original visibility weights (in this case without the application of {\tt statwt}), while the orange bars stem from the spatial scatter.
  \label{fig:Fvar_LB19}}
\end{figure}

\begin{figure}
  \centering
\includegraphics[width=0.8\columnwidth]{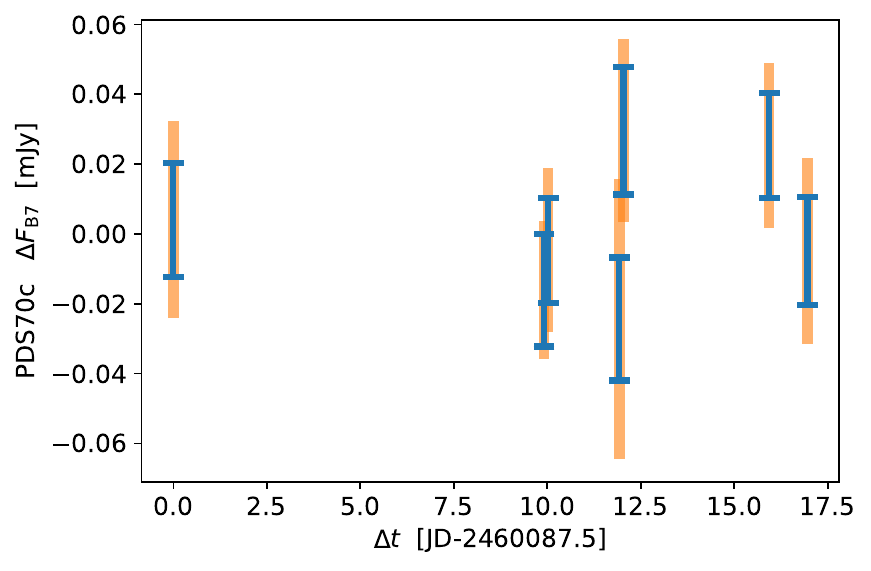}
\caption{Same as Fig.\,\ref{fig:Fvar}, but for the IB23 dataset
  (2023 Band\,7 observations). 
    \label{fig:Fvar_2023}}
\end{figure}

\begin{figure*}
  \centering
\includegraphics[width=\textwidth]{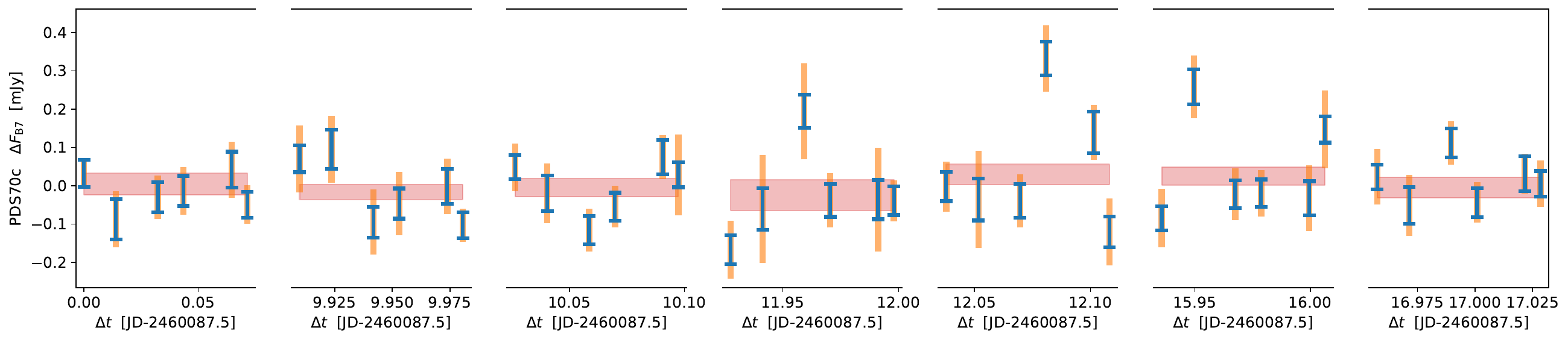}
  \caption{Same as Fig.\,\ref{fig:Fvar}, but for the fine-grained IB23 dataset, splitting each EB into 6 intervals. The blue error bars stem from  the  {\tt statwt} visibility weights, while the orange bars stem from the spatial scatter. The coarser measurements shown in transparent red boxes, with total vertical extension of 2$\sigma$, are reproduced from Fig.\,\ref{fig:Fvar_2023}, and correspond to extractions over the full EB. \label{fig:Fvar_2023_finegrained}}
\end{figure*}

We applied differential photometry to the LB19 Band\,7 dataset \citep[][]{Benisty2021ApJ...916L...2B}, which was acquired in 3
EBs, two from 27/7/2019, and one from 30/7/2019, and each
$\sim$\,1h15m long. Differential photometry yields the residual flux
densities $\Delta F_{\rm B7}$ shown in Fig.\,\ref{fig:Fvar_LB19}, where no
variability can be picked up among the three EBs.

We also revisited the full reduction and alignment of the IB23
dataset, and applied time-differential photometry for each EB, as
shown in Fig.\,\ref{fig:Fvar_2023}. The weighted-average residual flux
density is $\Delta F_{\rm B7} = 0.2\pm 8$\,$\mu$Jy, with a weighted
scatter of 18\,$\mu$Jy, which represents $\sim$15\% of the flux
density reported by \citep[][of
$121\pm13$\,$\mu$Jy]{Dominguez2025AA...702A..18D}.  The reduced
$\chi^2$ is 0.6 (with 6 degrees of freedom), and consistent with a
constant distribution. We therefore reach the same conclusion as
\citet[][]{Dominguez2025AA...702A..18D}, in the sense that the
observed scatter stems purely from the thermal uncertainties (which
appear to be somewhat conservative).

In an attempt to pick-up fine-grained variability, we split the seven EBs
from IB23 in six equal intervals, each $\sim$20\,min long, and aligned
the resulting ``sub-EBs'' using {\tt VisAlign}. The reference EB was
chosen as  2/6/2023, as it was acquired under the best
conditions. The corrections in flux scales were all less than 2\%, and
the pointing corrections less than 16\,mas, which reflects the
excellent weather. We then concatenated the aligned sub-EBs for joint
imaging, and applied differential photometry (i.e. we used the
residuals for the photometry of PDS\,70c in the $uv$-plane).

The fine-grained measurements of $\Delta F_{\rm
  B7}$ shown in Fig.\,\ref{fig:Fvar_2023_finegrained} seem to be
indicative of variability, as the reduced
$\chi^2$ is 1.67 (with 41 degrees of freedom). This corresponds to a
detection of variability at 99.6\% confidence, or
2.62\,$\sigma$. The weighted scatter is $\sigma_s =
13\pm2\,\mu$Jy, with bootstrapped uncertainties. The intrinsic
variability of PDS\,70c, $\sigma_{\rm
  int}$, can be estimated as an additional source of noise, added in
quadrature to each measurement of $\Delta F_{\rm
  B7}(t)$, such that the reduced
$\chi^2$ of the PDS\,70c time-series is brought down to unity. In this
case we find $\sigma_{\rm int} =
59\pm25\,\mu$Jy, with bootstrapped uncertainties. Fixing $F_{\rm
  B7}$ to the value for IB23 in Table\,\ref{table:PDS70cprev}, this
scatter corresponds to $49\%\pm21\%$.

However, the limitations on self-calibration discussed in
Sect.\,\ref{sec:residuals} suggest that the rapidly varying antenna
gains represent a source of systematic noise that cannot be corrected
through alignment to a global flux scale with {\tt VisAlign}. Also,
such non-thermal residuals would not be accounted for in the spatial
scatter along the orbit of PDS\,70c, since that region is
approximately devoid of signal. A test on synthetic and constant point
sources is required, with the same flux density as PDS\,70c, and at
the positions described in Sect. \ref{sec:diffphot}. These three
control point-sources have $\Delta F_{\rm B7}(t)$ time series which
result in reduced $\chi^2$ values of 1.35, 1.29 and 1.36, which are
all consistent with a constant value (the corresponding significance
of a non-constant value is $\sim 1.4\sigma$).

\end{appendix}
\end{document}